\documentclass[twocolumn,preprintnumbers,amsmath,amssymb]{revtex4}

\usepackage[english]{babel}
\usepackage{graphicx}
\usepackage{dcolumn}
\usepackage{bm}
\usepackage{color}

\newcommand{\tix}{\tilde{\chi}}
\newcommand{\beq}{\begin{equation}}
\newcommand{\eeq}{\end{equation}}
\newcommand{\beqr}{\begin{eqnarray}}
\newcommand{\eeqr}{\end{eqnarray}}
\newcommand{\trb}{\textup{Tr}_\textup{b}}
\newcommand{\trs}{\textup{Tr}_\textup{Q}}

\newcommand{\tr}{\textup{Tr}}

\begin{document}


%
\title{A consistent flow of entropy}
\author{Mohammad H. Ansari}
\affiliation{Kavli Institute of Nanoscience, Delft University of Technology, P.O. Box 5046, 2600 GA Delft, The Netherlands}


\begin{abstract}
A common approach to evaluate entropy in quantum systems is to solve a master-Bloch equation to determine density matrix and substitute it in entropy definition. However, this method has been recently understood to lack many energy correlators. The new correlators make entropy evaluation to be different from the substitution method described above. The reason for such complexity lies in the nonlinearity of entropy. In this paper we present a pedagogical approach to evaluate the new correlators and explain their contribution in the analysis. We show that the inherent nonlinearity in entropy makes the second law of thermodynamics to carry new terms associated to the new correlators. Our results show important new remarks on quantum black holes.  Our formalism reveals that the notion of  degeneracy of  states at the event horizon makes an indispensable deviation  from black hole entropy  in the leading order.
\end{abstract}
\maketitle

One of the most challenging problems in quantum physics is to fundamentally identify the statistical behaviours of information quantities. One of the central quantities in modern condensed matter is entropy, a nonlinear operator in density matrix $\hat{\rho}$, i.e. $S=-\textup{Tr}\hat{\rho}\ln \hat{\rho}$. One can consider  the unitary evolution of a closed quantum system and this indicates the quantity is a conserved measure, \cite{jaegerbook}. We can ask that what we will observe if we have access only to a small part of the closed system. Information which is initially encoded locally in an out-of-equilibrium state, as the system evolves in time, becomes encoded non-locally and less visible to an observer confined to the subsystem. Therefore one can assume there is flow of information out of the subsystem. From technical perspective an easier problem to solve is the special case of an open quantum system coupled to a large environment, \cite{kamenev}. The evolution of an open system using a Bloch-master equation and one can study evolution toward thermal equilibrium without explicit reference to the environment.

Quantum information (Shannon) theory can help us understand why the state of an open system might be expected to come close to the thermal Gibbs state. The free energy of the open system with density matrix $\rho$ is defined as $F(\rho)=\langle E(\rho)  \rangle  - S(\rho)/kT$ where $E(\rho)$ denotes the expectation value of the Hamiltonian in the Gibbs states. An open system in contact with a thermal reservoir will prefer the Gibbs state if it wishes to minimize its free energy. The second law of thermodynamics implies that the Gibbs state has the lowest possible free energy. Expressing the free energy $F(\rho_\beta)$ of the Gibbs state as  $F(\rho_\beta)=- \beta^{-1} \ln (\tr e^{-\beta H})$ indicates how the second law can be related to  the flow of relative entropy of $\rho$ and $\rho_\beta$: $S(\rho || \rho_\beta) = \tr (\rho \ln \rho) - \tr (\rho \ln \rho_\beta) = \beta (F(\rho)-F(\rho_\beta)) \geq 0 $, \cite{lieb}.

In quantum systems there are two major troubles with entropy: 1) its consistent evaluation, and 2) its measurement using physical quantities. The first problem goes back in its nonlinearity in density matrix. Entropy is commonly evaluated in open quantum systems using the  following substitution method: 1)  solving the non-unitary quantum evolution equation of open quantum system  to determine its reduced density matrix $\hat{\rho}_r(t)$ at arbitrary time $t$,  and 2) substituting $\hat{\rho}_r(t)$ in the definition of entropy $S=-\textup{Tr}\hat{\rho_r}\ln \hat{\rho_r}$.   

Recently it was understood that the  quantization of this quantity due to its nonlinearity is much more cumbersome such that using the substitution method gives rise to incomplete evaluation \cite{bej}. Following this study we noticed the existence of a large class of energy relaxations between open quantum system and its surrounding environment.\cite{Ansari15-1}  This can be more clarified when we represent entropy  in terms of the Renyi entropies $S_M= \textup{Tr}_r \{\rho_r\}^M$ in the limit of $S=-\lim_{M\to 1} dS_M/dM$.   For Gibbs states at temperature $T$ the Renyi entropies correspond to the difference of free energies at two temperatures, i.e., $\ln S_M = F(T)-F(T/M)$, \cite{{renyi}}. We developed a formalism to evaluate the time evolution of $(\rho_r)^M$ based on the Keldysh technique \cite{keldysh}. Our formulation shows that the consistent evaluation of entropy in open quantum system must take place using the following steps \cite{Ansari16}: 
\begin{enumerate}
\item computing the time evolution of the operator $(\hat{\rho}_r)^M$ using an extended Keldysh formalism in multiple parallel worlds\cite{an-jetp},  then
\item  substituting the solution in the definition of  Renyi entropy, and finally 
\item analytical continuing the result to the limit of $M\to 1$. 
\end{enumerate}

\begin{figure}
\begin{center}
a) \includegraphics[scale=0.25]{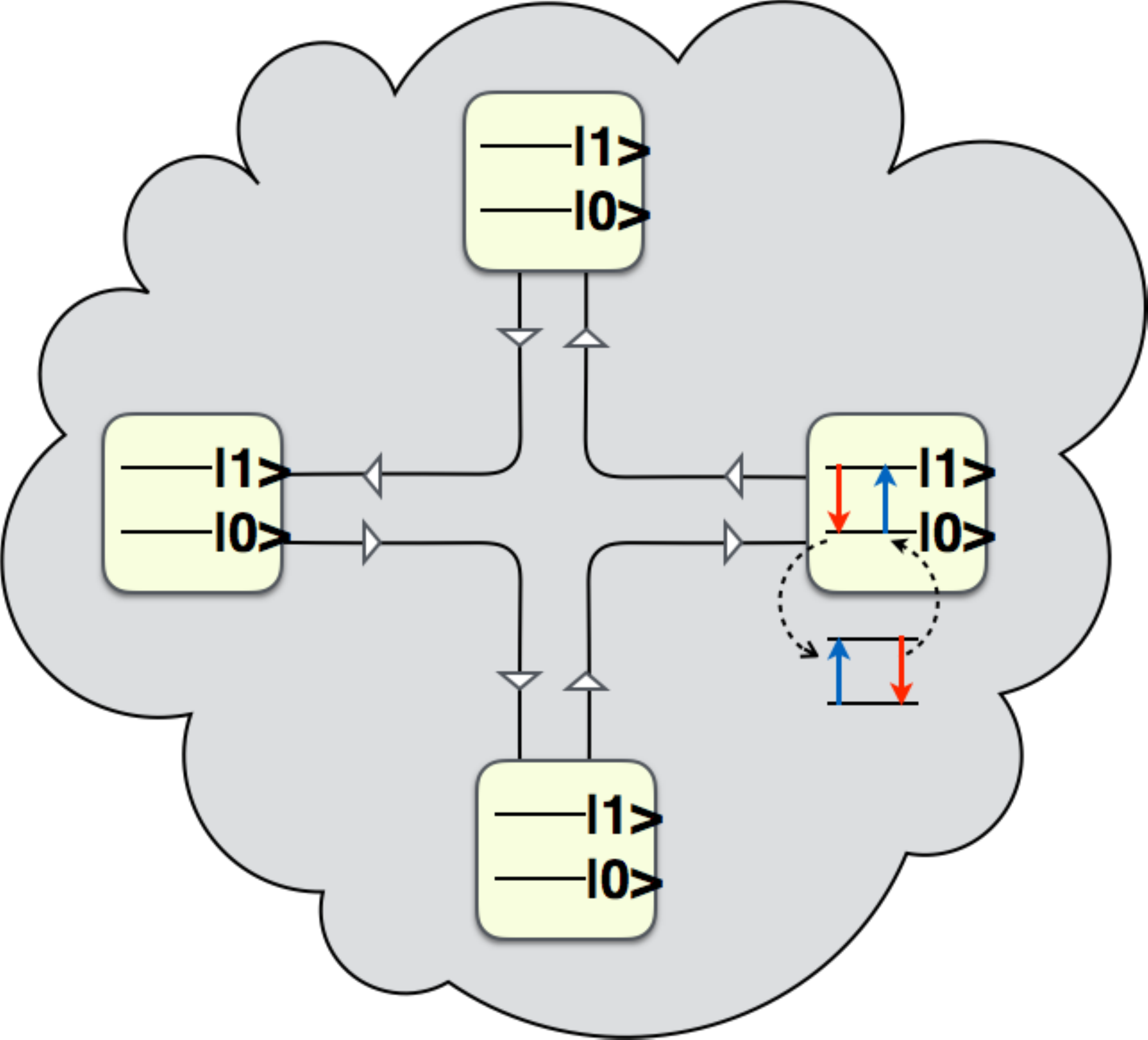}\\ b) \includegraphics[scale=0.25]{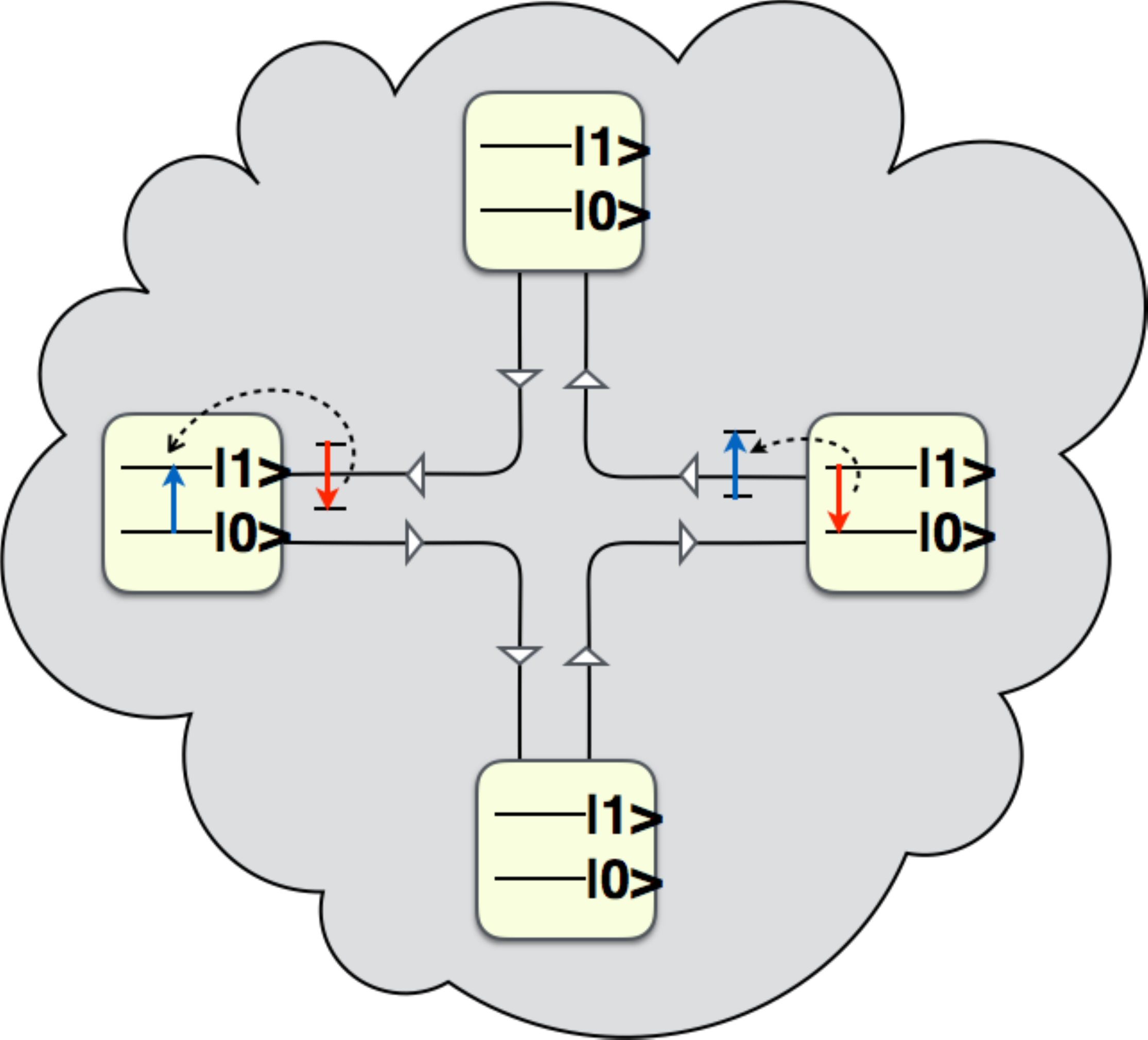}
\caption{(Color online) Schematic flow of Renyi entropy of degree $M=4$ between a two level system and its environment. The energy levels $|0,1\rangle$ are resonantly coupled to a large environment.  Each box denotes a copy of density matrix and the directed lines in between are the direction of exchanging information between density matrix copies. a) Incoherent flow, a qubit copy relaxes a photon to the environment and takes it back from environment. b) Coherent flow, the right qubit copy relaxes a photon to the environment and the photon after passively unchanging the upper copy of density matrix is returned to the leftmost copy of the density matrix. 
 }
\label{fig1}
\end{center}
\end{figure}

The entropy evaluated following these steps has two parts: incoherent and coherent, \cite{Ansari15-1}.  The incoherent part is equivalent to the heat dissipation in the form that is familiar to everyone through the second law of thermodynamics. The coherent part of the entropy flow is absent in the evaluation of entropy using substitution method and we showed it exist in nonlinear measures of density matrix where more than one density matrix to describe the interaction between system and environment. This new part simply describes how a photon created as a result of contact between system and environment in a copy of density matrix can be annihilated not in that density matrix but in another copy.  Fig. (\ref{fig1} a and b) schematically the incoherent and coherent flows of Renyi entropy of degree $M=4$ between a two level system (say a qubit) and its surrounding environment, respectively. The four copies of the qubit density matrix are indicated in boxes and placed in contact with a large environment.  Given that standard (single-world) correlator in equilibrium heat bath is proportional to the dynamical susceptibility using the so-called KMS (Kubo-Martin-Schwinger) relation \cite{kms},  the multiple-world  exchanges in the entropy flow between open quantum system and its environment take place using a new type of correlators, namely generalized KMS correlators \cite{Ansari15-1}. 

The second problem with entropy in quantum system. This quantity cannot be evaluated from immediate measurements, as its evaluation requires reinitialization of the density matrix between many successive measurements in a probe environment. The direct measurements of density matrix for a probe environment requires characterization of reduced density matrix of an infinite system, which is a rather nontrivial procedure and needs the complete and precise reinitialization of the initial density matrix.  An alternative approach to measure and control entropy exchange in open quantum systems is to find possible relation between entropy flow and measurable physical quantities. Recently we proved in Ref. \cite{Ansari15-2} that there is such an exact correspondence in the weak coupling limit. This makes possible to measure the new type of heat dissipation through coherent flows.  Similar relation has been worked out in electronic charge transfer by Levitov and Klich in Ref. \cite{Klich}. 

In this paper, we take pedagogical  steps to evaluate the flow of entropy in a simple quantum heat engine made of a two level quantum system in contact with heat reservoirs. For this aim we take the most simple atom-light interaction model. We will explicitly derive the generalized KMS correlators and using them we evaluate the consistent von Neumann entropy flow. Our result shows that steady state quantum coherences (i.e. off-diagonal elements of density matrix) in the second order perturbation theory influences the second law. Finally we discuss an application in quantum gravity and show how the new correlators modifies the Bekenstein-Hawking entropy of black holes.


%
\section{The Model}
\label{model}

Our aim in this section is to take  pedagogical steps to evaluate the Renyi entropy flow in the quantum heat engine that contains the simplest heat bath using the extended Keldysh formalism\cite{an-jetp}.  Although the abstract form of the entropy flow in a generic system can be found elsewhere\cite{{Ansari15-1},{Ansari16},{Ansari15-2}}, however the explicit form of entropy flow between a qubit and environment provide an opportunity to expect novel quantum phenomena emerging in quantum computation. This model will also motive a curious reader to study other types of baths and electron reservoirs.  

The Hamiltonian is $H=H_{0}+V$ with non-interacting qubit and photon reservoir part $H_0$,  
\beq  
\hat{H}_{Q}= E_0|0\rangle \langle 0| + E_1|1\rangle \langle 1|, \ \ \ \ 
\hat{H}_b = \sum_{\bf q} \hbar \omega_{\bf q} \hat{b}_{\bf q}^\dagger \hat{b}_{\bf q},  \label{eq.H0} 
\eeq with $\hat{b}_{{\bf q}}$ ($\hat{b}_{{\bf q}}^\dagger$) being annihilation (creation) photon operator with  momentum ${\bf q}$ in the reservoir. 
 
 The qubit is externally driven for example in cavity of fundamental frequency $\omega \approx \omega_Q=(E_1-E_0)/\hbar$.  The interaction Hamiltonian is 
\beq
 \hat{V}\left( t \right) =  \sum_{\bf q} \hbar\left\{ c_{\bf q}\left|0\rangle\langle 1\right|\hat{b}_{\bf q} \left(t\right)+ c_{\bf q}^* \left|1\rangle\langle 0\right|\hat{b}_{\bf q}^\dagger\left(t\right) \right\} ,
\label{eq.V}
\eeq
with the complex coupling energy $c_{\bf q}$. 

We assume adiabatic switching of the perturbation such that far in the past $t\to -\infty$ the coupling is absent, and the density matrix is the direct product of subsystems. The coupling slowly grows achieving actual values at a long time in the past of present time $t$. The time evolution of density matrix $\rho_{xy}$ in the quantum system formally takes place as 
\beq
\hat{\rho}(t)=Te^{i\int_{\infty }^t d\tau \hat{V}(\tau)}\hat{\rho}(-\infty) \bar{T}e^{i\int_{-\infty}^t d\tau \hat{V}(\tau)},
\label{rhot}
\eeq
with $T$ ($\bar{T}$) being (anti-) time ordering operator.  Using standard Keldysh formalism one can expand the operator in terms of $\hat{b}_{\bf q}$ and $\hat{b}^\dagger_{\bf q}$  operators in all orders; this sets the time ordering along the Keldysh contour that assumes opposite timing for bra and ket state evolutions. From the perturbative expansion in the second order of density matrix, one can see the density matrix of quantum system can be expressed using the correlator $S(\tau) =  \langle (\trs \hat{V}\left(t\right) ) (\trs \hat{V}\left(t+\tau\right)) \rangle $ being a photon exchange correlator and $\langle \cdots \rangle$ denotes trace over photon states re-summed over all such states, see Appendix \ref{app1} for details.    The integral for thermally equilibrium baths can be simplified to $\int_{0}^{\infty}S(\tau)  \exp(i\omega\tau) d\tau = (1/2) S\left(\omega\right)+i\Pi\left(\omega\right)$ with the Fourier transform of the correlator $S$ and $\Pi$ defined  as follows: 
\beqr \nonumber
&& {S\left(\omega\right)}  =  \hbar^2 \sum_{\bf q} 2\pi  |c_{\bf q}|^2 \left\{ \langle \hat{b}_{\bf q}\hat{b}^{\dagger}_{\bf q}\rangle \delta\left(\omega+\omega_{Q}\right)  \right. \\  && \left. \qquad \qquad \qquad \qquad \qquad + \langle\hat{b}^{\dagger}_{\bf q}\hat{b}_{\bf q}\rangle\delta\left(\omega-\omega_{Q}\right) \right\} \label{eq. P} \\ \nonumber
&& {\Pi\left(\omega\right)} = \hbar^2 \sum_{\bf q}  |c_{\bf q}|^2 \left\{ \langle \hat{b}_{\bf q}\hat{b}^{\dagger}_{\bf q}\rangle\left(\frac{1}{\omega+\omega_{Q}}\right)  \right. \\  && \left. \qquad \qquad \qquad \qquad \qquad  + \langle \hat{b}^{\dagger}_{\bf q} \hat{b}_{\bf q}\rangle\left(\frac{1}{\omega-\omega_{Q}}\right) \right\} 
\label{eq. Q}
\end{eqnarray}
The correlator $S(\omega)$ in thermally equilibrium heat baths is proportional to dynamical susceptibility of heat bath according to the Kubo-Martin-Schwinger (KMS) relation. Dynamical susceptibility in defined as $\chi\left(\omega\right)  =  (-i/\hbar) \int_{-\infty}^{0}  \langle [ \trs (\hat{V}\left(t+\tau\right)), \trs (\hat{V}\left(t\right)) ]\rangle \exp({-i\omega\tau})d\tau$. The KMS relation is $S= \hbar \bar{n}\tilde{\chi}  $ with  $\tilde{\chi}= (\chi-\chi^*)/i$ and $\bar{n}=1/(\exp(E/k_BT)-1)$ being the Bose distribution function. 

Evaluating R\'{e}nyi entropies requires time evolution of integer powers of density matrix.   Consider a closed system with total---namely  \emph{world}---density matrix $\rho$ made of two interacting subunits $A$ and $B$. The reduced density matrix for system $A$ is  $\rho_A=\textup{Tr}_{B} \rho$. The Renyi entropy of degree $M$ for the system $A$ is defined as $\ln S_M^{(A)}=\ln \textup{Tr}_A \{ \rho_A ^M\}$. If the two systems do not interact, the entropies are conserved $d\ln S_M^{(A,B)}/dt=0$; however for interacting heat baths in thermal equilibria, a steady flow of entropy is expected from one heat bath to another one. This is similar to the steady flow of charge in an electronic junction that connects two leads kept at different chemical potentials\cite{ansariqp}. Defining the Renyi entropy flow in $A$ as $-d\ln S_M^{(A)}/dt$,  there is a conservation law for  $d\ln S_M^{(A+B)}/dt$; however due to the inherent non-linearity $d (\ln S_M^{(A)}+\ln S_M^{(B)})/dt\neq 0$ and equality holds only approximately, subject to volume dependent terms.\cite{nazarov11}

For the evaluation of entropy flow $-d\ln S_M/dt$ in the second order perturbation we need to compute $d \rho^M/dt$. Considering the initial value $\rho_0$ of the density matrix $\rho(t) = \rho_0 +\rho^{(1)}(t)+O(2)$ with the first order $\rho^{(1)}(t) =- i \int_0^t dt' [H(t'),\rho(t)]$ and $d\rho(t)/dt = \delta^{(1)}(t) + \delta^{(2)}(t)+O(3)$ with the first order $\delta^{(1)}(t) =- i  [H(t),\rho(t)]$. The flow of nonlinear measure can be expanded as follows: $d\rho^M/dt=(d\rho/dt)\rho^{M-1}+\rho (d\rho/dt)\rho^{M-2}+\cdots+\rho^{M-1}(d\rho/dt)$. Using these perturbative theory one can conclude that
\beqr
\label{eq. 1} \nonumber
\frac{d\rho^M}{dt} &=& \left\{  \delta^{(2)} \rho_0^{M-1} +  \rho_0 \delta^{(2)} \rho_0^{M-2}+ \cdots+\rho_0^{M-1} \delta^{(2)}\right\} + \nonumber \\ && \nonumber 
 \bigg\{ \delta^{(1)} \left[ \rho^{(1)} \rho_0^{M-2} + \rho_0 \rho^{(1)} \rho_0^{M-3}+ \rho_0^2 \rho^{(1)} \rho_0^{M-4} + \cdots\right] \bigg. \\ \nonumber
&& \ \ \  + \rho_0 \delta^{(1)} \left[ \rho^{(1)} \rho_0^{M-3}  \rho_0 \rho^{(1)} \rho_0^{M-4}+\cdots\right]  \\ \nonumber
&& \ \ \  + \rho_0^2 \delta^{(1)} \left[ \rho^{(1)} \rho_0^{M-4}  \rho_0 \rho^{(1)} \rho_0^{M-5}+\cdots\right]   \\ && \ \ \    + \cdots  +    \bigg. \rho_0^{M-2} \delta^{(1)}  \rho^{(1)} \bigg\}  \eeqr
where the first line in Eq. (\ref{eq. 1}) represent single-world photon exchanges and the remaining are the exchange  of photon between different worlds.

\begin{figure}
\begin{center}
(a)\includegraphics[scale=0.35]{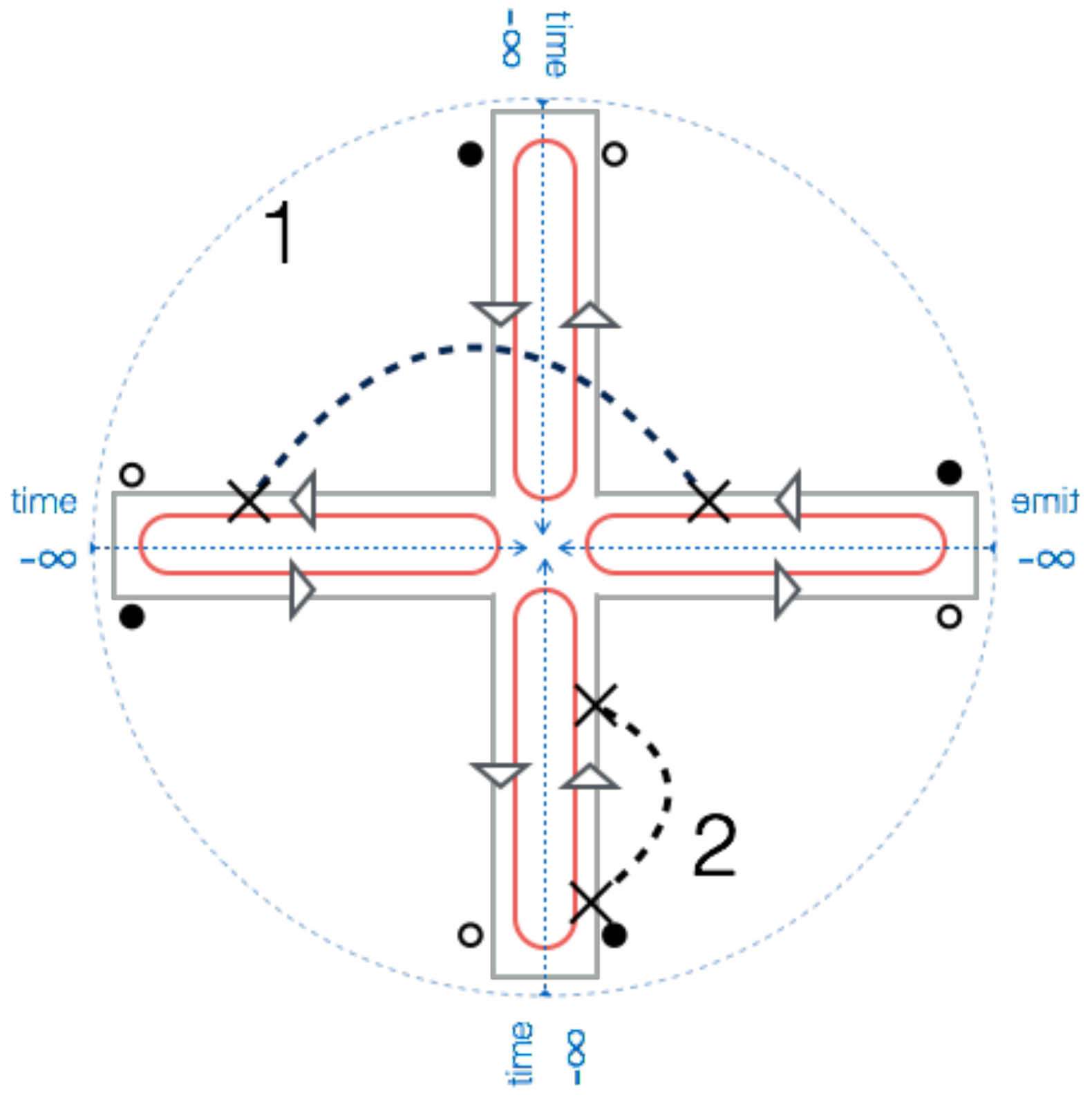}\\
(b) \includegraphics[scale=0.35]{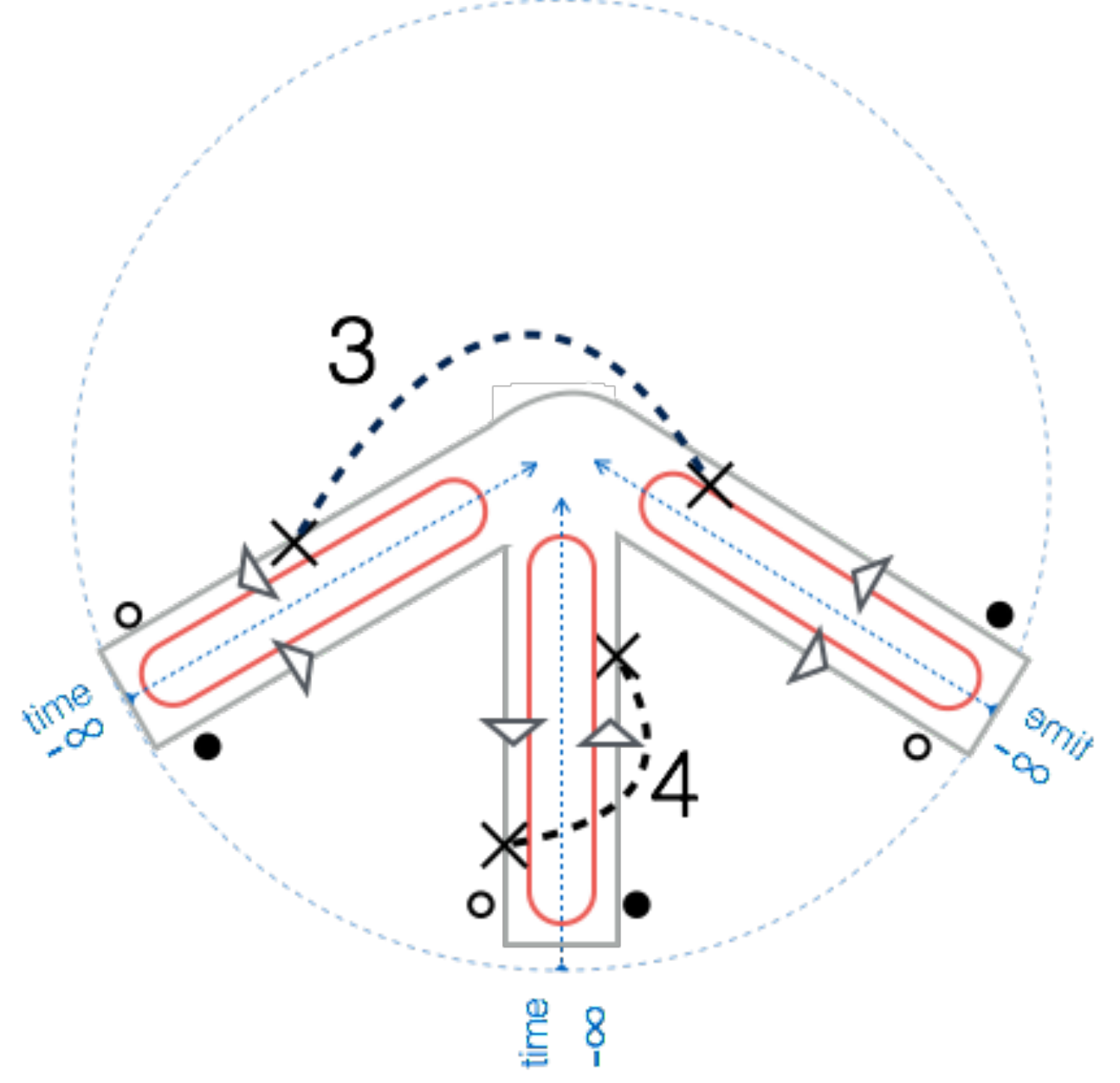} 
\caption{(Color online)  Two typical diagrams in the the calculation of Renyi entropy flow in system A for $M=3$. Time evolves from past at the perimeter toward present at the center. The contours for system A (B) are depicted as inner red (outer grey) lines. $M=4$ in (a) and  M=3 in (b). The interaction between A and B is denoted by a cross and dashed lines indicate correlators. The correlators (1,3) exchanges photons coherently between left and right worlds,  while the correlator (2,4) exchanges photon incoherently within one world. Black (while) circles denote ket (bra) states. Ket (bra) states evolve in the (opposite) direction of time flow. }
\label{diags}
\end{center}
\end{figure}

Computing the flow of Renyi entropy can take place using a diagrammatic representation of  Eq. (\ref{eq. 1}). We extended standard Keldysh formalism  in the time interval $(-\infty, t]$ in \cite{{an-jetp},{Ansari15-2}}, where we diagrammatically draw $M$ parallel worlds (each of which made of bra and ket-contours of a closed world). We used the term multiple parallel world for our formalism due to the specific preference in representing the worlds in parallel configuration. However, what is the most important concept in the diagrammatically representing the worlds is that the worlds can either passively transfer  photons coming from one worlds into another one, or they can actively exchange the energy between different subsystems. We choose a circular representation for the keldysh contours in multiple worlds. This,  although provide identical results but,  turns out to be rather helpful in computing other nonlinear measures such as  conditional and relative entropies. In Fig. (\ref{diags}) we perform the evaluation on a modified diagram that instead of drawing $M$ worlds in parallel, we draw them on the radius of a circle.  The circle perimeter denote time at $-\infty$ and the present time is at the center. As a contour moves toward the center the state it carries evolves forward in time. Each contour is a double line,  the internal one (red) for the evolution of A and the outer (grey) line for B. Figure (\ref{diags}a and b) denote $d \ln S_4^{(B)}/dt$ and $d \ln S_{3}^{(B)}/dt$, respectively. Notice that since these diagrams compute the flow of entropy in system B, the out (grey) contours encompass all worlds.

Let us consider $A$ being a thermal heat bath at temperature $T=1/k_B\beta$ with diagonal density matrix on the eigenmodes $|\{n_{i}\}\rangle\equiv|n_{1},\cdots,n_{q},\cdots\rangle$
with probability $\mathcal{P}\left(\left\{ n_{i}\right\} \right)=\prod_{i}p\left(n_{i}\omega_{i}\right)$
with $p\left(n_{i}\omega_{i}\right)=\exp\left(-\beta\hbar\omega_{i}n_{i}\right)/Z\left(\omega_{i}\right)$ and $Z(\omega_i)$ being the partition function of mode $\omega_i$. Consider the creation  operator of a photon $\hat{b}_{\bf q}^\dagger$ to take place somewhere on the contour that encompasses all worlds. This changes the state of heat bath as long as the created photon is not annihilated. The annihilation $\hat{b}_{\bf q}$ occurs after the photon kinematically passes through $N$ worlds. The generalized correlator  turn out to have the following form
\beq
S^{N,M}(\tau)=\frac{\trb\bigg\{ \trs (\hat{V}(t) )\ \rho_\textup{b}^N \ \trs (\hat{V}(t+\tau)) \ \rho_\textup{b}^{M-N} \bigg\}}{\trb (\rho_\textup{b}^M )}
\label{eq.corr}
\eeq
with a typical internal correlator that by substituting the explicit density matrix of heat bath can be simplified to

\beqr 
\frac{ \left\langle \hat{b}_{\bf q}\rho_{\textrm{bath}}^{n} \hat{b}_{\bf q}^{\dagger}\rho_{\textrm{bath}}^{M-n} \right\rangle }{ \left\langle \rho_{\textrm{bath}}^{M} \right\rangle }   & = & \frac{\exp({\hbar\omega_{Q}(M-n)/k_BT})}{\exp({\hbar\omega_{Q}M}/k_BT)-1}
\label{eq: corr b rho bdag rho}
\eeqr

Using the definition of $\hat{V}$ in Eq. (\ref{eq.V}) and the generalized correlator becomes
\beqr \nonumber
&&S^{N,M}\left(\omega\right)  =  \nonumber 2\pi\hbar^2 |c|^2  \\  &&  \times \left\{ \frac{e^{\beta\hbar\omega_{Q}\left(M-N\right)}}{e^{\beta\hbar\omega_{Q}M}-1}\delta\left(\omega+\omega_{Q}\right) +\frac{e^{\beta\hbar\omega_{Q}N}}{e^{\beta\hbar\omega_{Q}M}-1}\delta\left(\omega-\omega_{Q}\right)\right\} \nonumber \\
\label{eq: S(w,N)}
\eeqr
which satisfies  the following `generalized detailed balance' relation between absoption and emission:  $S^{n,M}\left(-\omega\right)  =  S^{M-n,M}\left(\omega\right) $. One can also show that these correlators satisfy a `generalized KMS relation' with the bath dynamical susceptibility $\tilde{\chi}$ against perturbation, 
\beq
 S^{N,M}\left(\omega\right)= \hbar  \bar{N}(M\omega /T)   \exp({\hbar N\omega }/k_BT)  \tilde{\chi}\left(\omega\right)
\label{eq. KMS}
\eeq

Similar to the what mentioned above, see the paragraph after Eq. (\ref{rhot}),   time evolution of the operator $\rho(t)^M$ between $(-\infty,t]$ can be determined through the evaluation of $\int_0^\infty d\tau \left\langle X\left(t\right)\rho_b^{N} X\left(t+\tau\right) \rho_b^{M-N} \right\rangle e^{i\omega\tau}/{\left\langle \rho_{\textrm{env}}^{M} \right\rangle }$, which is $(1/2) S^{N,M}(\omega) + i \Pi^{N,M}(\omega)$, given the definition of  $ \Pi^{N,M} = (1/2\pi) \int dz  S^{N,M}(z) / (\omega-z)$. Detailed analysis of evaluating these new correlators and the dynamical susceptibility can be found in Appendices (\ref{app1}) and  (\ref{app2}).

We implement the extended Keldysh formalism for the analysis of Renyi entropy flow, defined as $ \mathcal{F}_M= -{d \ln S_M}/{dt}$. Detailed analysis following Eq. (\ref{eq. 1})  and using the generalized KMS relation will result the following total flow of Renyi entropy:
\beqr 
\mathcal{F}_M  & =&  \frac{M  \bar{n}(M\omega_Q) }{\hbar \omega_Q\ \bar{n}(\omega_Q) \bar{n}((M-1)\omega_Q)}  \left\{   \langle E \rangle -   Q^{(c)} \right\} 
   \label{eq. rflow gen}
\eeqr
with 
\beqr 
\langle E \rangle    &=&    \Gamma_{\downarrow}  p_1  \omega_Q -  \Gamma_{\uparrow} p_0   \omega_Q  \\ 
Q^{(c)} &=&      2 \rho_{10}\rho_{01} \omega_Q      \label{eq. Qs}
\eeqr
and the emission  rate $ \Gamma_{\downarrow} = (\bar{n}(\omega_Q) +1) \tilde{\chi} $ and absorption rate $ \Gamma_{\uparrow} =  \bar{n}(\omega_Q) \tilde{\chi} $. 
In Eq. (\ref{eq. rflow gen}) there are two types of flows contributing: (i) the incoherent flow  $\langle E \rangle$, for quantum leaps on energy levels, and (ii) the coherent flow $Q^{(c)}$ for the exchange of energy through the quantum coherence. 

Consequently, the consistent von Neumann entropy flow $\mathcal{F}_S$ in an open quantum system does not follow the textbook formulation $dS/dt   \neq  { \langle E \rangle  }/{T}  $, instead our consistent formalism turn out to have the following structure \beq   
 \frac{d S}{dt}   =  \frac{ \langle E \rangle  }{T}- \frac{ Q^{(c)}}{T},  
   \label{eq. vn entropy}
\eeq

We show that in a resonantly driven qubit coupled to environment the von Neumann entropy is not directly related to the decoherence of qubit in the way that textbook second law of thermodynamics suggests. Instead, there is a non-trivial part of information flow between qubit and environment that can be represented, which here we called it the coherent flow of entropy and this amount should be subtracted from the flow caused by energy dissipation.  

\section{Remarks on black hole entropy}

The second law of thermodynamics admits that black holes have finite entropy flow \cite{Hawking}. Following the idea Bekenstein conjectured \cite{beken} that the black hole entropy was proportional to the area of its event horizon divided by the Planck area. The idea is that a black hole event horizon is made of $N$ number of discrete area patches. Each patch of area $a_o$ may carry  minimal information of 2 states. The event horizon of area $A=Na_o$ will carry  the total number of $2^N$ configurations and therefore carries the entropy $S\sim\ln 2^N= k A$ with $k$ being $\ln 2/a_o$. 

Semiclassical quantization analysis predicts a discrete area for rotating uncharged holes, charged holes, Kerr and extremal Kerr holes, and some others, (for the list of references see \cite{Ansari10} and references therein). Loop quantum gravity also supports the discreteness on any surface \cite{asht}. Because the area of a black hole surface is proportional to energy in a black hole, the black hole mass is likely to be quantized as well, although the quantization has not be justified independently. 

Black hole mass decreases when it radiates; therefore its quantum of mass decreases by a finite value after one emission, similar to the way atoms decay.  We showed that the complete spectrum of area carries an inherent degeneracy that scales  power law with the eigenvalue of area \cite{Ansari07-2}. This degeneracy leads to a visible quantum effect in the spectroscopy of radiation from a quantum black hole \cite{Ansari07-1}. In a different approach, based on isolated horizon theory, a similar-yet-of-different-nature degeneracy was associated to the horizon  area states \cite{ABCK}.  These two pictures provides a deeper mathematics for the above-mentioned entry of black holes from counting its microscopic states.

In counting the microscopic states of a black hole even horizon, it is non-trivially assumed that the off-diagonal elements of density matrix for each patch of area is zero. However, this assumption has not a strong basis as the quantum evolution of black hole states is unknown. In fact resonantly driving  a two level system coupled to environment  leads to double degenerate quantum states coupled to environment. We take an advantage from the analogy and consider a simple model of black hole that carries a steady quantum coherence. 

The second law of thermodynamics on black holes between black hole mass $M$ and its even horizon area $A$, i.e. $ A=  16 \pi M^2 $, provides a link between the flow of black hole mass $\delta M$  and the flow of its event horizon area  $\delta A$ in the following form ${\delta M}= ({1}/{ 32 \pi  M})  {\delta A}$. We can consider that the area $\delta A$  carries degeneracy.  The energy corresponding to $\delta M$ is exchanged between the hole and its environment. Therefore the simple model described above is applicable as a simple model for quantizing black holes.  According to the laws on black hole:
\beq
\label{eq. bh} 
\frac{d M}{dt}= \frac{1}{ 32 \pi  M}  \frac{d A}{dt}
\eeq

By substituting Eq. (\ref{eq. vn entropy}) in Eq. (\ref{eq. bh}) and given the relation between energy transfer and mass change $d M = d E$ and the black hole temperature $T=1/8\pi k M$,  one concludes that 

  \beq
\label{eq. bh2} 
 \frac{d S}{dt}   = \frac{ 1   }{ 4 }  \frac{dA}{dt} -   {8\pi   M}
Q^{(c)} \eeq

In the simple case of assuming the exchanged energy is $\omega_o=\delta M$ the relation is simplified to $dS/dt=(1/4)(1-2|\rho_{01}|^2)dA/dt$. This is crucially important result. In all previous works on black hole physics it was claimed that the only correction to the Bekenstein-Hawking entropy is in a log-term correction. However, our consistent evaluation of entropy that takes care of the nonlinearity of entropy clearly indicates that the presence of quantum coherent flow of entropy in the leading term that can make the total entropy flow prefactor  $1-2|\rho_{01}|^2$ much less than one.  Contemplating on this new result helps to further understand the seemingly information paradox in black holes. 

%
%

\section{Summary}
\label{discussion}

We calculated the Renyi entropy flow of bath-system entanglement in an open quantum system. To achieve a full dynamical quantization of entropy we use the extended Keldysh formalism in multiple ordered worlds.   Our exact calculation shows that quantum coherence directly influences the flow of  von Neumann  and Renyi entropies.  We also discussed how this result can lead to deviation form the second law of thermodynamics. As a simple application we recalculated the black hole von Neumann entropy as a result of the presence of degenerate area eigen states on the even horizon. We show that our consistent formulation of von Neumann entropy explicitly show that the entropy of a quantum black hole does not satisfy the Bekenstein-Hawking law. 

\section{Acknowledgements}
The research leading to these results has received funding from the European Union Seventh Framework Programme (FP7/2007-2013) under grant agreement n° 308850 (INFERNOS).

\appendix

\section{Multi-world  correlators}
\label{app1}

Using the interaction Hamiltonian defined in Eq. (\ref{eq.V}) we can evaluate the correlator integral using two functions $P$ and $\Pi$, 
\beq
\int_{0}^{\infty}\langle V\left(t\right)  V\left(t+\tau\right)\rangle e^{i\omega\tau}d\tau =  P\left(\omega\right)+iQ\left(\omega\right)
\label{eq. forward prop}
\eeq
with $\langle \cdots \rangle$ indicating trace over environment density matrix and 
\beqr \nonumber
P\left(\omega\right) & \equiv & \hbar^2\pi |c|^{2} \sum_{q}\langle b_{q}b_{q}^{\dagger}\rangle\delta\left(\omega+\omega_{q}\right)   + \langle b_{q}^{\dagger}b_{q}\rangle\delta\left(\omega-\omega_{q}\right) \label{eq. P1}\\  \nonumber
\Pi\left(\omega\right) & \equiv & \hbar^2 |c|^2 \sum_{q}\langle b_{q}b_{q}^{\dagger}\rangle\left(\frac{1}{\omega+\omega_{q}}\right) + \langle b_{q}^{\dagger}b_{q}\rangle\left(\frac{1}{\omega-\omega_{q}}\right) 
\label{eq. Q1}
\end{eqnarray}

From the detail form of the functions $P$  in Eqs. (\ref{eq. Q1}), one can show that for negative frequency the following relations hold
\begin{eqnarray}\nonumber
&&\frac{P\left(-\omega\right)}{e^{\beta\hbar\omega}}  =   |c|^2\left[\pi\sum_{q}\frac{1}{e^{\beta\hbar\omega_{q}}-1}\delta\left(\omega-\omega_{q}\right)\right. \\ \nonumber && \qquad \left.\qquad \qquad +\frac{e^{\beta\hbar\omega_{q}}}{e^{\beta\hbar\omega_{q}}-1}\delta\left(\omega+\omega_{q}\right)\right] = P\left(\omega\right)
 \label{eq. pPi}
\end{eqnarray}

From the definitions in Eq. (\ref{eq. pPi})  it is easy to show that $\Pi$ is not independent function and in fact for all frequencies it can be obtained as follows:
 
\beqr
\Pi(\omega)&=&\frac{1}{2\pi} \int d\nu  \frac{S(\nu)}{\omega-\nu}
\label{eq. Q S}\\
\Pi(-\omega)&=&-\frac{1}{2\pi} \int d\nu  \frac{S(\nu)e^{\beta \nu}}{\omega-\nu}
\label{eq. Q S neg w}
\eeqr

Moreover, the Fourier transform of the correlator $S(\tau)$, defined below  (\ref{rhot}),  is 

\begin{eqnarray}\nonumber
S\left(\omega\right) & \equiv & \int_{-\infty}^{\infty}\langle V\left(t\right)V\left(t+\tau\right)\rangle e^{i\omega\tau}d\tau\\
 & = & 2P\left(\omega\right)
 \label{eq. S and P}
\end{eqnarray}

  \section{Dynamical susceptibility}
\label{app2}

Generalized dynamical susceptibility is defined below and can be simplified using the $P$ and $\Pi$ functions:

\begin{eqnarray} \nonumber
\chi\left(\omega\right) & = & -\frac{i}{\hbar}\int_{-\infty}^{0}\langle\left[V\left(t+\tau\right), V\left(t\right)\right]\rangle e^{-i\omega\tau}d\tau\\ \nonumber
 & = & \frac{1}{\hbar}\left(\Pi\left(\omega\right)+\Pi\left(-\omega\right) \right)  -\frac{i}{\hbar}\left[P\left(\omega\right)-P\left(-\omega\right)\right]
 \label{eq. chi general}
\end{eqnarray}

From above definitions  and after a few lines of simple algebra one can show
\beqr
\chi(\omega)=\frac{1}{2\pi\hbar} \int dz \frac{S(z)}{\bar{n}(z) (\omega-z)} + \frac{i  S(\omega)}{2\hbar \bar{n}(\omega)} 
\label{eq. chi S and int S}
\eeqr

Determining spectral density in terms of susceptibility enables us to simplify the dynamics of density matrix. From eq. (\ref{eq. chi general})  we can split the dynamical susceptibility into symmetric $\chi^+$ and asymmetric $\chi^-$ parts 
\beq
\chi(\omega)=\chi^+(\omega)+i\tilde{\chi}(\omega)
\label{eq. chi to chip and chitilde}
\eeq
where $2\chi^{\pm}(\omega)\equiv \chi(\omega) \pm \chi(-\omega)$ and $\tilde{\chi}\equiv -i \chi^-$. The following two identities determine spectral density and $\chi^+$ in terms of $\tilde{\chi}$:
\beqr
S(\omega)&=& 2\hbar \bar{n}(\omega)
 \tilde{\chi}(\omega)  
 \label{eq. S tilde chi}\\
\chi^+(\omega) &=& \frac{1}{\pi} \int dz  \frac{
 \tilde{\chi}(z)}{\omega-z} 
 \label{eq. chiplus tilde chi}
\eeqr
and since $\chi(\omega)^*=\chi(-\omega)$ one can show that 

\beq 
\tilde{\chi}(\omega)= \frac{\chi(\omega)-\chi^*(\omega)}{2i}.
\label{eq. tildechi to chi and chistar}
\eeq

Also in general from eqs. (\ref{eq. pPi}) and (\ref{eq. S tilde chi}) it is simple to prove

\beq
\tix(-\omega)=-\tix(\omega)
\eeq

Putting the relations in Eq. (\ref{eq. P1}) and (\ref{eq. chi general}) one can see that $\tix$ consists of a continuous series of $\delta$-peaks
\beqr \nonumber
\tilde{\chi}(\omega)&=&\hbar \pi |c|^2 \sum_q \left[ \delta(\omega- \omega_q)    -\delta(\omega+\omega_q) \right]
\label{eq. tildechi delta dirac}
\eeqr

Simple algebra shows  $\left[\tilde{\chi}(\omega)\right]^*=\tilde{\chi}(\omega)$,  indicates the invariance of Eqs. (\ref{eq. S tilde chi}) and (\ref{eq. chiplus tilde chi}) under complex conjugate.

\end{document}